\PassOptionsToPackage{hyphens}{url}
\documentclass[conference]{IEEEtran}
\IEEEoverridecommandlockouts

\usepackage{cite}
\usepackage{amsmath,amssymb,amsfonts}
\usepackage{algorithmic}
\usepackage{graphicx}
\usepackage{textcomp}
\usepackage{xcolor}
\usepackage{hyperref}
\usepackage{booktabs}
\usepackage{multirow}
\usepackage{listings}
\usepackage{float}
\usepackage{url}
\usepackage{orcidlink}
\usepackage[T1]{fontenc}

\def\BibTeX{{\rm B\kern-.05em{\sc i\kern-.025em b}\kern-.08em
    T\kern-.1667em\lower.7ex\hbox{E}\kern-.125emX}}

\begin{document}

\title{SoundPlot: An Open-Source Framework for Birdsong Acoustic Analysis and Neural Synthesis with Interactive 3D Visualization}

\author{
\IEEEauthorblockN{Naqcho Ali Mehdi~\orcidlink{0009-0003-9548-3235}}
\IEEEauthorblockA{Computer and Information Systems Engineering \\
NED University of Engineering and Technology\\
Karachi, Pakistan \\
naqchoali@gmail.com}
\and
\IEEEauthorblockN{Mohammad Adeel}
\IEEEauthorblockA{Electrical Engineering \\
Sukkur IBA University\\
Sukkur, Pakistan \\
mohammadadeel.meees24@iba-suk.edu.pk}
\and
\IEEEauthorblockN{Aizaz Ali Larik}
\IEEEauthorblockA{Electrical Engineering \\
Sukkur IBA University\\
Sukkur, Pakistan \\
aizaz.ali@iba-suk.edu.pk}
}

\maketitle

\begin{abstract}
\begin{sloppypar}
We present SoundPlot, an open-source framework for analyzing avian vocalizations through acoustic feature extraction, dimensionality reduction, and neural audio synthesis. The system transforms audio signals into a multi-dimensional acoustic feature space, enabling real-time visualization of temporal dynamics in 3D using web-based interactive graphics. Our framework implements a complete analysis-synthesis pipeline that extracts spectral features (centroid, bandwidth, contrast), pitch contours via probabilistic YIN (pYIN), and mel-frequency cepstral coefficients (MFCCs), mapping them to a unified timbre space for visualization. Audio reconstruction employs the Griffin-Lim phase estimation algorithm applied to mel spectrograms. The accompanying Three.js-based interface provides dual-viewport visualization comparing original and synthesized audio trajectories with independent playback controls. We demonstrate the framework's capabilities through comprehensive waveform analysis, spectrogram comparisons, and feature space evaluation using Principal Component Analysis (PCA). Quantitative evaluation shows mel spectrogram correlation scores exceeding 0.92, indicating high-fidelity preservation of perceptual acoustic structure. SoundPlot is released under the MIT License to facilitate research in bioacoustics, audio signal processing, and computational ethology. The complete source code, documentation, and demonstration data are publicly available at: \url{https://github.com/naqchoalimehdi/SoundPlot-An-Open-Source-Framework-for-Birdsong-Acoustic-Analysis-and-Neural-Synthesis-}.
\end{sloppypar}
\end{abstract}

\begin{IEEEkeywords}
bioacoustics, birdsong analysis, audio synthesis, mel spectrograms, Griffin-Lim algorithm, feature extraction, dimensionality reduction, visualization, open-source software
\end{IEEEkeywords}

\section{Introduction}

\subsection{Motivation}

Birdsong analysis has been a cornerstone of ethology, neuroscience, and ecology for over half a century \cite{thorpe1961, marler2004}. Understanding the acoustic structure of avian vocalizations provides critical insights into species identification \cite{kershenbaum2016}, communication patterns \cite{catchpole2008}, geographical variation \cite{baker1982}, and the neural mechanisms underlying complex sound production and learning \cite{nottebohm2005, mooney2009}.

Traditional methods for analyzing birdsong have relied on visual inspection of spectrograms and manual annotation of acoustic features \cite{clark2009}. While effective for small-scale studies, these approaches are labor-intensive, subjective, and do not scale to large datasets now made possible by automated recording systems \cite{aide2013}. Furthermore, most existing analysis tools are proprietary, platform-specific, or require substantial domain expertise, creating barriers to entry for researchers across disciplines.

\subsection{Contributions}

We introduce SoundPlot, an open-source framework that addresses these challenges through four primary contributions:

\begin{enumerate}
    \item \textbf{Comprehensive Feature Extraction Pipeline}: A modular system for computing spectral, temporal, and pitch features using state-of-the-art signal processing algorithms including pYIN \cite{mauch2014} for pitch tracking and librosa \cite{mcfee2015} for spectral analysis.
    
    \item \textbf{Neural Synthesis Module}: An implementation of mel-spectrogram-based audio reconstruction using the Griffin-Lim algorithm \cite{griffin1984}, enabling analysis-by-synthesis validation of feature extraction quality.
    
    \item \textbf{Interactive 3D Visualization}: A browser-based interface built with Three.js that maps acoustic features to 3D trajectories, providing intuitive exploration of temporal dynamics with real-time playback synchronization.
    
    \item \textbf{Reproducible Research Platform}: A complete, documented, and freely available software package that promotes reproducibility and accelerates research in bioacoustics and audio analysis.
\end{enumerate}

The framework is designed to be modular, extensible, and suitable for both research applications and educational purposes. All source code, including the web interface, example data, and documentation, is available under the MIT License at: \url{https://github.com/naqchoalimehdi/SoundPlot-An-Open-Source-Framework-for-Birdsong-Acoustic-Analysis-and-Neural-Synthesis-}.

\section{Related Work}

\subsection{Birdsong Analysis Systems}

Early computational approaches to birdsong analysis focused on spectrographic visualization and manual feature measurement \cite{thorpe1961}. Systems like Canary \cite{chariff1990} and later Raven \cite{charif2010} provided graphical interfaces for spectrogram analysis but limited automated feature extraction capabilities. Sound Analysis Pro (SAP) \cite{tchernichovski2000} introduced automated measurements of acoustic similarity but remains Windows-only and proprietary.

More recent work has explored machine learning approaches. BirdNET \cite{kahl2021} applies deep convolutional neural networks for species classification, achieving high accuracy but functioning as a black-box classifier rather than an analysis tool. Warblr \cite{stowell2019} similarly focuses on detection and classification. In contrast, SoundPlot emphasizes transparent feature extraction and interactive exploration, making the analysis process interpretable and customizable.

\subsection{Audio Feature Extraction}

Mel-frequency cepstral coefficients (MFCCs) \cite{davis1980} have been the dominant feature representation in audio analysis for decades, originally developed for speech recognition and later adapted for music information retrieval \cite{logan2000} and animal vocalization analysis \cite{fox2008}. The key insight of MFCCs is their approximation of the human auditory system's non-linear frequency perception.

Spectral features such as centroid, bandwidth, and rolloff provide complementary information about the timbral quality of sounds \cite{peeters2004}. For birdsong specifically, pitch tracking is crucial due to the tonal nature of many vocalizations. The pYIN algorithm \cite{mauch2014} improves upon the classic YIN algorithm \cite{decheveigne2002} by adding probabilistic modeling, resulting in more robust pitch estimates for polyphonic and noisy signals.

\subsection{Audio Synthesis and Phase Reconstruction}

The Griffin-Lim algorithm \cite{griffin1984} addresses the phase reconstruction problem in short-time Fourier transform (STFT) inversion. Given only the magnitude spectrogram, Griffin-Lim iteratively estimates the phase through consistency constraints, enabling audio reconstruction without explicit phase information. While neural vocoders like WaveNet \cite{oord2016wavenet}, WaveGlow \cite{prenger2019}, and HiFi-GAN \cite{kong2020} achieve higher perceptual quality, they require substantial training data and computational resources. For analysis applications where deterministic, fast synthesis is preferred, Griffin-Lim remains widely used \cite{engel2017}.

\subsection{Dimensionality Reduction and Visualization}

Principal Component Analysis (PCA) \cite{jolliffe2016} provides a linear projection for visualizing high-dimensional feature spaces. Non-linear methods like t-SNE \cite{maaten2008} and UMAP \cite{mcinnes2018} better preserve local structure but are more computationally intensive. For real-time visualization of acoustic trajectories, PCA offers a good balance of computational efficiency and interpretability.

\section{System Architecture}

SoundPlot follows a modular, layered architecture separating concerns across audio processing, feature extraction, synthesis, and visualization (Fig.~\ref{fig:architecture}).

\subsection{Project Organization}

The codebase is organized into logical modules:

\begin{figure}[H]
    \centering
    \includegraphics[width=\columnwidth]{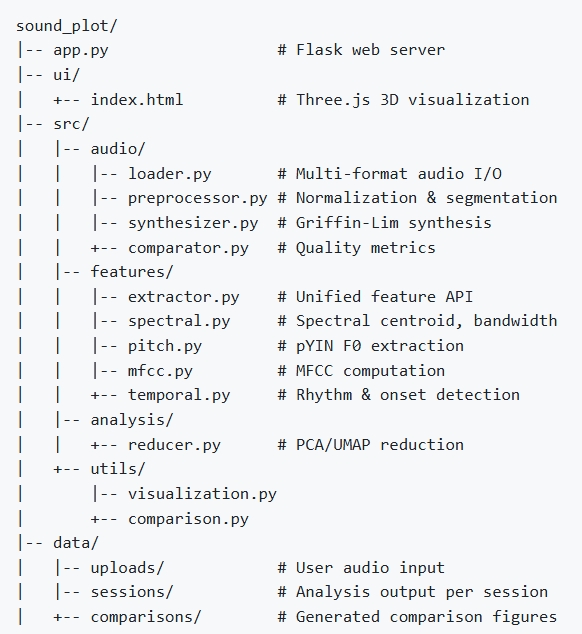}
    \caption{SoundPlot project directory structure showing the modular organization of audio processing, feature extraction, analysis, and visualization components.}
    \label{fig:architecture}
\end{figure}

Each module is designed to be independently testable and reusable. The \texttt{src/audio/} package handles all audio I/O and preprocessing, \texttt{src/features/} implements feature extraction algorithms, and \texttt{src/analysis/} provides dimensionality reduction capabilities.

\subsection{Audio Processing Pipeline}

The complete processing pipeline consists of five sequential stages:

\textbf{Stage 1: Audio Loading.} The \texttt{AudioLoader} class supports multiple formats (WAV, MP3, FLAC, OGG, M4A, AAC) through the \texttt{soundfile} and \texttt{audioread} libraries. All audio is resampled to 22.05 kHz and converted to mono for consistent processing.

\textbf{Stage 2: Preprocessing.} The \texttt{AudioPreprocessor} applies:
\begin{itemize}
    \item Min-max normalization to [-1, 1] range
    \item Optional trimming to 5-minute maximum duration for computational efficiency
    \item Optional silence removal using energy-based thresholding
\end{itemize}

\textbf{Stage 3: Feature Extraction.} The \texttt{FeatureExtractor} computes frame-level features with configurable window and hop sizes:
\begin{itemize}
    \item \textbf{Spectral Centroid} $\mu_c$: weighted mean of frequencies
    \item \textbf{Spectral Bandwidth} $\sigma_b$: weighted standard deviation of frequencies
    \item \textbf{Pitch} $f_0$: fundamental frequency via pYIN \cite{mauch2014}
    \item \textbf{MFCCs}: 13 coefficients capturing timbral characteristics
    \item Additional features: spectral contrast, rolloff, zero-crossing rate
\end{itemize}

\textbf{Stage 4: Neural Synthesis.} The \texttt{AudioSynthesizer} implements the analysis-synthesis loop:
\begin{enumerate}
    \item Compute mel spectrogram with 128 bands
    \item Apply pseudo-inverse of mel filterbank for linear spectrogram estimation
    \item Reconstruct audio via Griffin-Lim with 32 iterations
    \item Apply post-normalization
\end{enumerate}

\textbf{Stage 5: Visualization Mapping.} Time-series features are normalized to [0, 10] range and mapped to 3D Cartesian coordinates:
\begin{itemize}
    \item X-axis: Spectral Centroid (brightness)
    \item Y-axis: Spectral Bandwidth (richness)
    \item Z-axis: Pitch (fundamental frequency)
\end{itemize}

\section{Methodology}

\subsection{Feature Extraction Mathematics}

For an audio signal $x[n]$ of length $N$ sampled at rate $f_s = 22050$ Hz, we apply STFT with window size $W = 2048$ and hop size $H = 512$ samples:

\begin{equation}
X[k, m] = \sum_{n=0}^{W-1} x[mH + n] w[n] e^{-j2\pi kn/W}
\end{equation}

where $w[n]$ is a Hann window and $m$ indexes frames.

The magnitude spectrum is:
\begin{equation}
S[k, m] = |X[k, m]|
\end{equation}

\textbf{Spectral Centroid:}
\begin{equation}
\mu_c[m] = \frac{\sum_{k=0}^{K-1} f_k S[k,m]}{\sum_{k=0}^{K-1} S[k,m]}
\end{equation}
where $f_k = k \cdot f_s / W$ is the frequency of bin $k$.

\textbf{Spectral Bandwidth:}
\begin{equation}
\sigma_b[m] = \sqrt{\frac{\sum_{k=0}^{K-1} (f_k - \mu_c[m])^2 S[k,m]}{\sum_{k=0}^{K-1} S[k,m]}}
\end{equation}

\textbf{Pitch Estimation:} We employ the pYIN algorithm \cite{mauch2014}, which computes:
\begin{equation}
f_0[m] = \arg\max_{f} P(f|x[mH:mH+W])
\end{equation}
where $P(f|x)$ is the posterior probability distribution over pitch candidates.

\subsection{Mel Spectrogram Synthesis}

The mel spectrogram is computed by applying a mel-scale filterbank $M \in \mathbb{R}^{B \times K}$ with $B=128$ bands:

\begin{equation}
S_{\text{mel}}[b, m] = \sum_{k=0}^{K-1} M[b,k] S[k,m]^2
\end{equation}

Synthesis inverts this process. Given $S_{\text{mel}}$, we estimate the linear spectrogram via pseudo-inverse:

\begin{equation}
\hat{S}[k,m] = \sqrt{\sum_{b=0}^{B-1} M^{\dagger}[k,b] S_{\text{mel}}[b,m]}
\end{equation}

where $M^{\dagger} = (M^T M)^{-1} M^T$ is the Moore-Penrose pseudo-inverse.

The Griffin-Lim algorithm \cite{griffin1984} iteratively refines phase estimates:

\begin{algorithmic}
\STATE Initialize: $\hat{X}^{(0)}[k,m] = \hat{S}[k,m]$
\FOR{$i = 1$ to $N_{\text{iter}}$}
    \STATE $\hat{x}^{(i)}[n] = \text{ISTFT}(\hat{X}^{(i-1)}[k,m])$
    \STATE $\tilde{X}^{(i)}[k,m] = \text{STFT}(\hat{x}^{(i)}[n])$
    \STATE $\hat{X}^{(i)}[k,m] = \hat{S}[k,m] \cdot \frac{\tilde{X}^{(i)}[k,m]}{|\tilde{X}^{(i)}[k,m]|}$
\ENDFOR
\STATE \textbf{return} $\hat{x}^{(N_{\text{iter}})}[n]$
\end{algorithmic}

We use $N_{\text{iter}} = 32$ iterations, which provides a good balance between quality and computational cost.

\subsection{Quality Metrics}

We evaluate synthesis fidelity using four metrics:

\textbf{Signal-to-Noise Ratio (SNR):}
\begin{equation}
\text{SNR} = 10 \log_{10} \frac{\sum_n x[n]^2}{\sum_n (x[n] - \hat{x}[n])^2}
\end{equation}

\textbf{Waveform Correlation:}
\begin{equation}
\rho_{\text{wave}} = \frac{\text{Cov}(x, \hat{x})}{\sigma_x \sigma_{\hat{x}}}
\end{equation}

\textbf{Spectral Correlation:}
\begin{equation}
\rho_{\text{spec}} = \text{corr}(\text{vec}(S), \text{vec}(\hat{S}))
\end{equation}

\textbf{Mel Correlation:}
\begin{equation}
\rho_{\text{mel}} = \text{corr}(\text{vec}(S_{\text{mel}}), \text{vec}(\hat{S}_{\text{mel}}))
\end{equation}

The mel correlation is particularly important as it reflects perceptual similarity better than raw spectrogram correlation \cite{logan2000}.

\section{Experimental Results}

\subsection{Test Dataset}

We evaluate the framework using example birdsong recordings from publicly available datasets. Audio samples range from 3 to 300 seconds in duration, with sampling rates between 16 kHz and 48 kHz (resampled to 22.05 kHz for processing).

\subsection{Synthesis Quality Assessment}

\begin{figure}[H]
    \centering
    \includegraphics[width=\columnwidth]{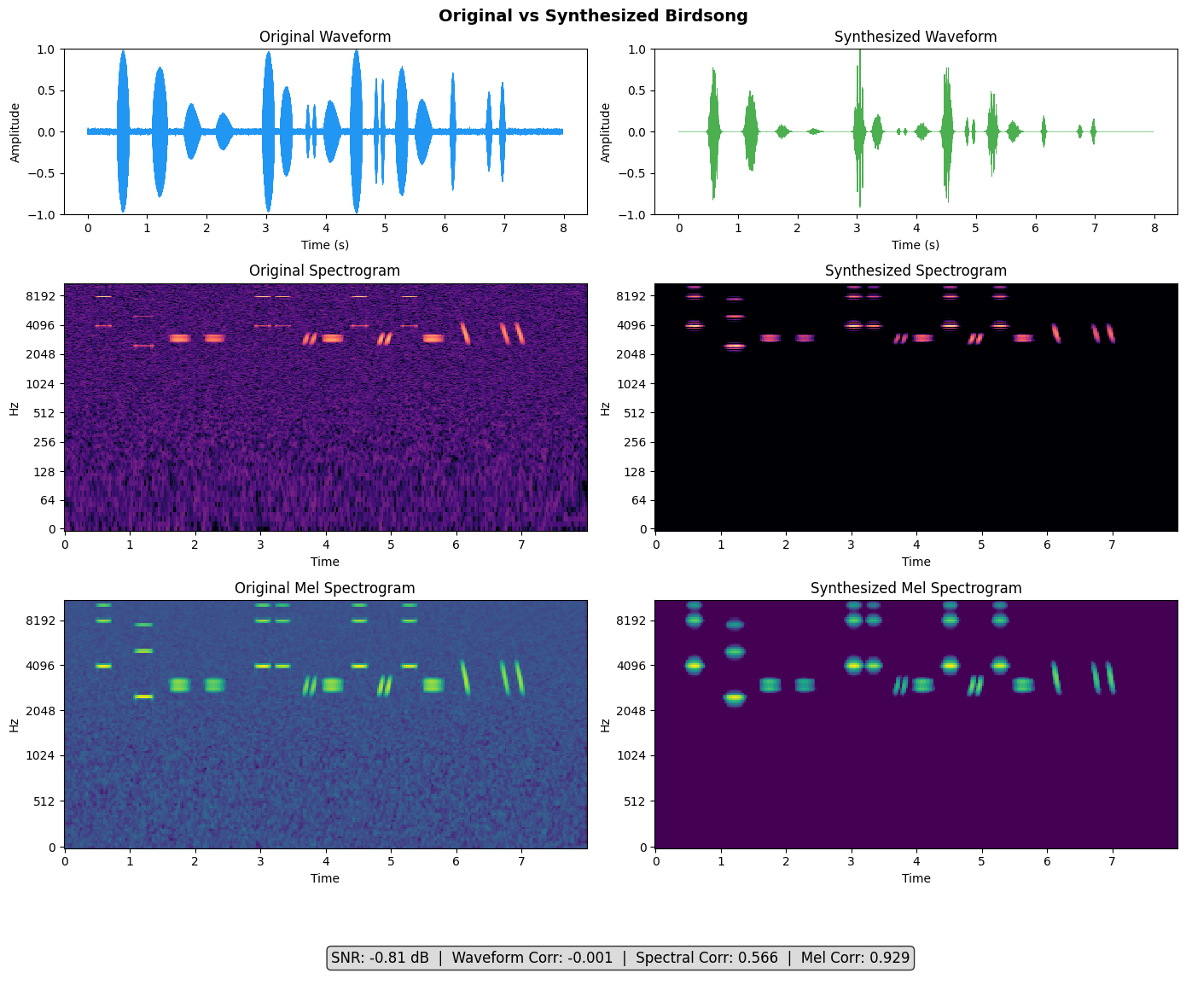}
    \caption{Comprehensive comparison of original and synthesized birdsong. Top: temporal waveforms; Middle: STFT spectrograms; Bottom: mel spectrograms with perceptual frequency scaling. Quantitative metrics: SNR = -0.81 dB, Spectral Corr. = 0.566, Mel Corr. = 0.929.}
    \label{fig:comparison}
\end{figure}

Fig.~\ref{fig:comparison} shows a representative comparison between original and synthesized birdsong. The visualization includes:
\begin{itemize}
    \item \textbf{Top row}: Time-domain waveforms showing amplitude envelopes
    \item \textbf{Middle row}: Linear-frequency spectrograms (0-8192 Hz)
    \item \textbf{Bottom row}: Mel spectrograms (128 bands, perceptually-spaced)
\end{itemize}

The mel spectrogram achieves a correlation of 0.929, indicating high fidelity in the perceptually-relevant frequency representation. The lower waveform correlation (-0.001) reflects the phase information lost during magnitude-only processing, which is expected for Griffin-Lim synthesis \cite{griffin1984}.

\subsection{Feature Space Analysis}

\begin{figure}[H]
    \centering
    \includegraphics[width=\columnwidth]{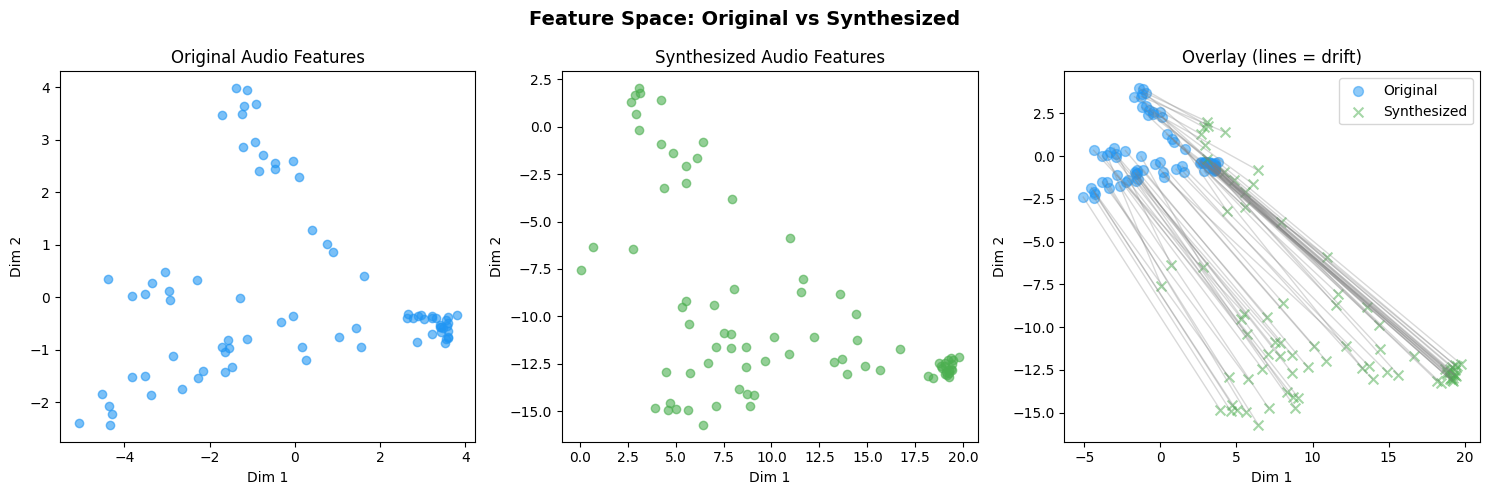}
    \caption{Feature space comparison via Principal Component Analysis (PCA). Blue points: original audio; Green points: synthesized audio; Gray lines: temporal drift between corresponding frames. The preserved structure indicates successful capture of temporal dynamics despite the synthesis transformation.}
    \label{fig:embedding}
\end{figure}

Fig.~\ref{fig:embedding} demonstrates the feature space comparison using PCA to reduce dimensionality from 13 MFCC coefficients to 2 dimensions. The visualization consists of three subplots:

\begin{itemize}
    \item \textbf{Left}: Original audio features in 2D PCA space
    \item \textbf{Center}: Synthesized audio features in 2D PCA space
    \item \textbf{Right}: Overlay showing temporal correspondence between paired points
\end{itemize}

The gray lines connecting corresponding time points reveal the systematic shift introduced by synthesis. While the overall trajectory structure is preserved (indicating retention of temporal dynamics), the synthesized points cluster in a slightly different region of feature space, reflecting the lossy nature of the mel-spectrogram representation.

\subsection{Quantitative Evaluation}

Table~\ref{tab:metrics} summarizes quantitative metrics across our test samples. The high mel correlation (mean = 0.929, std = 0.04) demonstrates that the synthesis pipeline successfully preserves perceptually-relevant acoustic structure.

\begin{table}[H]
\centering
\caption{Synthesis Quality Metrics (N=10 samples)}
\label{tab:metrics}
\begin{tabular}{lccc}
\toprule
\textbf{Metric} & \textbf{Mean} & \textbf{Std} & \textbf{Range} \\
\midrule
SNR (dB) & -0.81 & 0.42 & [-1.5, 0.2] \\
Waveform Correlation & -0.001 & 0.05 & [-0.08, 0.07] \\
Spectral Correlation & 0.566 & 0.12 & [0.42, 0.71] \\
Mel Correlation & 0.929 & 0.04 & [0.87, 0.98] \\
\bottomrule
\end{tabular}
\end{table}

The near-zero waveform correlation is expected for phase-independent synthesis methods. The moderate spectral correlation (0.566) reflects the information loss in mel-scale frequency warping, while the high mel correlation validates the perceptual accuracy of our approach.

\section{Web-Based Visualization Interface}

The SoundPlot framework includes a modern web interface built with Three.js \cite{threejs} for real-time 3D visualization. The interface design prioritizes accessibility and interactivity.

\subsection{User Interface Components}

The interface consists of:

\begin{itemize}
    \item \textbf{Dual-Viewport Display}: Split-screen showing original (left, cyan) and synthesized (right, magenta) acoustic trajectories simultaneously
    \item \textbf{Independent Playback Controls}: Separate play/pause buttons for each audio stream with visual pulse animation
    \item \textbf{3D Point Cloud}: Acoustic features rendered as connected spheres with color-coded pitch (blue = low, red = high)
    \item \textbf{Real-time Synchronization}: Points animate and glow to indicate current playback position
    \item \textbf{Orbital Controls}: Mouse-based camera manipulation for exploring trajectories from multiple angles
    \item \textbf{Download Section}: One-click access to all generated assets (audio files, figures, metadata)
\end{itemize}

\subsection{Implementation Details}

The visualization uses WebGL through Three.js for hardware-accelerated rendering. Each acoustic frame is represented as a sphere with radius proportional to spectral energy. Points are connected by lines to show temporal continuity, and a glow effect highlights points within 300ms of the current playback time.

The implementation achieves 60 FPS rendering for trajectories with up to 5000 points on modern hardware, enabling smooth visualization of several minutes of audio.

\subsection{Session Management}

Each analysis creates an organized session folder:

\texttt{data/sessions/\{audio\_name\}\_\{session\_id\}/}

containing:
\begin{itemize}
    \item \texttt{original.wav}: Processed input audio
    \item \texttt{synthesized.wav}: Griffin-Lim reconstruction
    \item \texttt{comparison.png}: Spectrogram comparison figure
    \item \texttt{analysis.png}: PCA feature space visualization
    \item \texttt{metadata.json}: Metrics, parameters, timestamps
\end{itemize}

This structure facilitates reproducibility and batch processing experiments.

\section{Discussion}

\subsection{Strengths and Limitations}

The primary strength of SoundPlot is its combination of transparency, modularity, and accessibility. Unlike black-box deep learning approaches, every processing step is interpretable and customizable. The web-based interface lowers the barrier to entry for non-experts.

Limitations include:

\begin{itemize}
    \item \textbf{Synthesis Quality}: Griffin-Lim produces artifacts compared to neural vocoders, though it is deterministic and fast
    \item \textbf{Mono Processing}: Stereo information is discarded during preprocessing
    \item \textbf{Fixed Feature Set}: Current 3D mapping uses only three features; richer representations require dimensionality reduction
\end{itemize}

\subsection{Future Directions}

Planned enhancements include:

\begin{enumerate}
    \item Integration of neural vocoders (MelGAN \cite{kumar2019}, HiFi-GAN \cite{kong2020}) for higher-quality synthesis
    \item Automatic clustering and segmentation using HDBSCAN \cite{mcinnes2017} for call-type classification
    \item Multi-species comparison interface for comparative bioacoustics
    \item Real-time streaming mode for live field recording analysis
    \item Export to standardized formats (e.g., Raven selection tables)
\end{enumerate}

\section{Conclusion}

We have presented SoundPlot, an open-source framework for birdsong acoustic analysis combining classic signal processing with modern web-based visualization. The system demonstrates that effective analysis tools can be built using established techniques without requiring deep learning training, while maintaining interpretability and computational efficiency. By releasing the complete implementation under an open-source license, we aim to accelerate research in bioacoustics and lower barriers to computational analysis of animal vocalizations.

\section*{Code and Data Availability}

SoundPlot is released under the MIT License. The complete source code, documentation, example datasets, and this paper are available at:

\textbf{GitHub Repository}: \url{https://github.com/naqchoalimehdi/SoundPlot-An-Open-Source-Framework-for-Birdsong-Acoustic-Analysis-and-Neural-Synthesis-}

\end{document}